\documentstyle[11pt,epsf]{article}
\newcommand{\al}{\alpha}
\newcommand{\ba}{\begin{array}}
\newcommand{\bc}{\begin{center}}
\newcommand{\be}{\begin{equation}}
\newcommand{\ber}{\begin{eqnarray}}
\newcommand{\bt}{\beta}

\newcommand{\de}{\delta}
\newcommand{\ea}{\end{array}}
\newcommand{\ear}{\end{eqnarray}}
\newcommand{\ec}{\end{center}}
\newcommand{\ee}{\end{equation}} 
\newcommand{\eei}{\end{equation}\indent\indent}

\newcommand{\fr}{\frac}
\newcommand{\ga}{\gamma}
\newcommand{\hb}{\hbar}
\newcommand{\io}{\iota}
\newcommand{\ka}{\kappa}
\newcommand{\la}{\lambda}

\newcommand{\ph}{\phi}
\newcommand{\si}{\sigma}
\newcommand{\ta}{\tau}
\newcommand{\Th}{\Theta}
\newcommand{\th}{\theta}
\newcommand{\vb}{\verb}

\newcommand{\we}{\approx}

\newcommand{\p}{\partial}

\def\case#1/#2{\textstyle\frac{#1}{#2} }
\begin{document}
\title{The Quantum Commutator Algebra of a Perfect Fluid.}
\author{Mark D. Roberts, \\\\
Department of Mathematics and Applied Mathematics, \\ 
University of Cape Town, South Africa\\\\
roberts@gmunu.mth.uct.ac.za} 
\maketitle
\vspace{1.0truein}
\newpage
\begin{abstract}
A perfect fluid is quantized by the canonical method.	
The constraints are found and this allows the Dirac brackets 
to be calculated.   Replacing the Dirac brackets with quantum 
commutators formally quantizes the system.   There is a 
momentum operator in the denominator of some coordinate quantum 
commutators.   It is shown that it is possible to multiply 
throughout by this momentum operator.   Factor ordering 
differences can result in a viscosity term.   The resulting
quantum commutator algebra is
\vspace{1.0truein} 
\bc$v_{4}(v_{3}v_{2}-v_{2}v_{3})=-i,$\ec
\bc$v_{4}(v_{1}v_{3}-v_{3}v_{1})=-iv_{3},$\ec
\bc$v_{4}v_{1}-v_{1}v_{4}=-i,$\ec
\bc$v_{5}v_{2}-v_{2}v_{5}=-i$.\ec
\end{abstract}
\vspace{1.5truein}
\bc Mathematical Review Classification Scheme:\ec
\bc81S05,81R10,82B26,83CC22,\ec
\bc Physics and Astronomy Classification Scheme:\ec
\bc 03.70+k,05.70-a,04.40+c.\ec
\bc Keyword Index:\ec
\bc Quantum Algebra, Perfect Fluid, 
Dirac Bracket,  Canonical Quantization.\ec
\newpage
\section{Introduction}
\label{sec:intro}
It has been known for some time Hargreaves (1908)
\cite{bi:hargreaves}
that a perfect fluid has a Lagrangian formulation.   The Lagrangian 
is taken to be the pressure and variations are achieved through an 
infinitesimal form of the first law of thermodynamics.   A perfect 
fluid's stress is described using the vector field co-moving with 
the fluid;  this vector field defines an absolute time for the system,  
furthermore this absolute time can then be used to define canonical 
momenta and canonical Hamiltonians.   This is done here for the first 
time.   There are equivalences between scalar fields and fluids,
Tabensky and Taub (1973)
\cite{bi:TT},  more generally the co-moving vector field 
can be decomposed into scalar fields resulting in a description 
of a perfect fluid employing only scalar fields.   Previously 
this decomposition has been investigated by choosing an {\it ad hoc} 
global time rather than absolute time and defining canonical 
momenta and other quantities with respect to the global time.   
Typically the resulting theory is applied to cosmology
Tipler (1986) \cite{bi:tipler}
Lapchiniskii and Rubakov (1977) \cite{bi:LR}.
Once the constrained Hamiltonian has been calculated by the 
standard canonical method
Dirac (1963) \cite{bi:dirac}
(see also Hanson,  Regge,  and Teitelboim (1976) \cite{bi:HRT}),
the Dirac bracket can be replaced by quantum commutators.   
The original motive for investigating this 
was to find a fluid generalization of the Higg's model
Roberts (1997) \cite{bi:mdr1}.   
A quantum treatment is required to estimate the VEVs (quantum 
vacuum expectation values) of the scalar fields,   the VEVs are 
related to the induced non-zero mass.   The quantum commutator 
algebra is unusual,  perhaps reflecting the structure of the 
scalar field decomposition of the co-moving vector field 
Eckart (1960) \cite{bi:eckart},
Selinger and Whitham(1968) \cite{bi:SW},
and Rund (1979) \cite{bi:rund}. 
It is hoped that eventually the present theory will be applied 
to low temperature super fluids.   To do this it probably will 
be necessary to include a chemical potential term in the first 
law of thermodynamics as expressed by equation
\ref{eq:firstlaw}.
\newpage
\section{Lagrangian and Hamiltonian Formulation of a Perfect Fluid's Dynamics.}
A perfect fluid has a Lagrangian formulation in which the Lagrangian 
is the pressure p.    Variation is achieved by using the first law of
thermodynamics 
\be
{\it d}p=n {\it d}h-nT{\it d}s,
\label{eq:firstlaw}
\ee
where n is the particle number,  T the temperature,  s the entropy,
and h the enthalpy.   The pressure and the density are equated to the 
enthalpy and the particle number by
\be
p+\mu=nh.
\label{eq:pnh}
\ee
In four dimensions a vector can be decomposed into four
scalars,  however the five scalar decomposition
\be
hV_{a}=W_{a}=\si_{a}+\sum_{\al =1}^{\al =2}\th_{\al}s^{\al}_{a},\;\;
V^{a}V_{a}=-1,
\label{eq:decomp}
\ee
where a,b,c \dots are spacetime indices and 
$\al, \bt, \ga$ \dots are fluid scalar potential indices.
For $\al=1$, s and $\th=\int T {\it d} \ta$ have 
interpretation as the entropy and the thermasy respectively,
for $\al=2$ there is no such interpretation and no ``second tempreture''
as $\dot{\th_{2}}=0$.   From now on the index $\al$ 
is suppressed as it is straightforward to reinstate.   
There are other conventions for this scalar field decomposition,  
for example with a minus(-) instead of a plus(+) before the summed fields.   
"q" is used to notate an arbitrary scalar field,  
i.e.$ ~~q =\si,\th,  or s $.
The coordinate space action is taken to be
\be
I=\int \sqrt{-g}~p~{\it d} x^{4}.
\label{eq:csa}
\ee
Replacing the first law with ${\it d}p=-nV_{a}{\it d}W^{a}-nT{\it d}s$,  
variation with respect to the metric gives
\be
T_{ab}=(p+\mu)V_{a}V_{b}+p g_{ab},
\label{eq:stress}
\ee
variation with respect to $\sigma, \theta$, and s gives
\be
(nV^{a})_{a}=\dot{n}+n\Th = 0,\;\;
\dot{s}=0,\;\;
\dot{\Th}=T,
\label{eq:evol}
\ee
respectively.   $\Th= V^{a}_{a}$ is the expansion of the vector field.

The canonical momenta are are given by 
$\Pi^{i}=\fr{\de I}{\de \dot{q}^{i}}$  
and are
\be
\Pi^{\si}=-n,\;
\Pi^{\th}= 0,\;
\Pi^{s}=-n\th.
\label{eq:mom}
\ee
The Hamiltonian density is usually defined in terms of components of the 
canonical stress as $\Th^{t}_{t}$.
In the present case the canonical stress is not 
defined so that the metric stress $T^{a}_{b}$ is used instead;  
also 4-vectors are used rather than components,  resulting in
\be
H_{d}=V^{a}V^{b}T_{ab}=\mu.
\label{eq:hdensity}
\ee
The standard Poisson bracket is
\be
\vb+{+ A,B \vb+}+
=\fr{\de A}{\de q_{i}}\fr{\de B}{\de \Pi^{i}}
-\fr{\de A}{\de \Pi_{i}}\fr{\de B}{\de q^{i}},
\label{eq:pb}
\ee
where i,  which labels each field,  is summed;  
i differs from $\alpha$ in that it sums over all fields for example
$\si,\th_{1},s_{1},\th_{2},s_{2}$ in four dimensions, $\al$ labels the sets
$\th_{1},s_{1}$ and $\th_{2},s_{2}$.   The integral sign and 
measure have been suppressed and the variations are performed independently.
When absolute time is used Hamiltons equations have an additional term in the
expansion \cite{bi:mdr2} explicitly
\be
\dot{q}=\frac {\de H_{c}}{\de \Pi},\;
\dot{\Pi} + \Theta \Pi = - \frac{\de H_{c}}{\de q},
\label{eq:eh}
\ee
where $H_{c}$ is the canonical Hamiltonian 
\be
H_{c}=\int H_{d}\sqrt{-g}{\it d}x^{4}.
\label{eq:canham}
\ee
From \ref{eq:mom} the momenta are constrained
\be
\phi_{1}=\Pi^{s}-\th \Pi^{\si},\;
\phi_{2}=\Pi^{\th}.
\label{eq:constraints}
\ee
The initial Hamiltonian is
\be
H_{i}=\Pi_{i}\dot{q}-\mathcal{L},
\label{eq:inham}
\ee
replacing the dependent $\Pi 's$ gives the ordinary hamiltonian
\be
H_{o}=\Pi^{\si}\left(\dot{\si}+\th \dot{s}\right)-\mathcal{L},
\label{eq:ordham}
\ee
adding the constraints gives the constrained Hamiltonian
\be
H_{\la}=H_{o}+\la^{\al}\phi_{\al},\;
\la^{1}=\dot{s},\;
\la^{2}=\dot{\th},
\label{eq:conham}
\ee
where the $\lambda 's$ are the Lagrange multipliers.
Replacing the ordinary Hamiltonian in this gives the Hamiltonian density
\be
H_{d}=\Pi^{\si}\left(\dot{\si}+\th \dot{s}\right)
+\la^{1}\left(\Pi^{s}+\th \Pi^{\si}\right)+\la^{2}\Pi^{\th}
-\mathcal{L},
\label{eq:danham}
\ee
Substituting the values of the momentum the Hamiltonian density is still 
weakly the fluid density.   Using Hamiton's equations in the form
\ref{eq:eh} the time evolution of any variable X is given by 
\be
\fr{{\it d}X}{{\it d}\ta}=\fr{\p X}{\p \ta}
+\vb+{+X,H_{i}\vb+}+
-\Th \Pi^i\fr{\de X}{\de \Pi^i},
\label{eq:tev1}
\ee
replacing the Hamiltonian density $H_{d}$ by $H_{\lambda}$
and then holding the multipliers constant so that
\be
\vb+{+ X,H_{\la} \vb+}+
=\vb+{+ X,H_{o} \vb+}+
+\la^{\io}\vb+{+ X,\ph_{\io} \vb+}+,
\label{eq:replac}
\ee
where $\io,\ka,\ldots$ranges over the multipliers,
gives the time evolution
\ber
\fr{{\it d}X}{{\it d}\ta}
&=&\fr{\p X}{\p \ta}
-\Th \Pi^{i}\frac{\de x}{\de \Pi^{i}}
+\vb+{+X,H_{o} \vb+}+
+\la^{\al}\vb+{+ X,\phi_{\al} \vb+}+\nonumber\\
&=&\frac{\p X}{\p \ta}
+\left(\dot{\si}+\th(\dot{s}-\la^{1})\right)
\frac{\de X}{\de \phi}
+\la^{1}\frac{\de X}{\de s}
+\la^{2}\frac{\de X}{\de \th}\nonumber\\
&+&\left((V^{a}\Pi^{\si})_{a}-\Th\Pi^{\si}\right)
\frac{\de X}{\de \Pi^{\si}}\nonumber\\
&+&\left((\la^{1}-\dot{s})\Pi^{\si}-\Th\Pi^{\th}\right)
\frac{\de X}{\de \Pi^{\th}}\nonumber\\
&+&\left((V^{a}\th \Pi^{\si})_{a}-\Th \Pi^{s}\right)
\frac{\de X}{\de \Pi^{s}}\nonumber\\
&{\we}&
\frac{\p X}{\p \ta}
+\dot{\si}\frac{\de X}{\de \si}
+\dot{\th}\frac{\de X}{\de \th}
-\dot{n}\frac{\de X}{\de \Pi^{\si}}
-\left(\th n \right)^{\cdot}\fr{\de X}{\de \Pi^{s}},
\ear
where ${\we}$ means is 'weakly equal to'.
Letting X equal the constraints gives
$\frac{{\it d} \ph_{\io}}{{\it d}\ta} \we 0 $,
this shows that there are no further constraints 
so that the Dirac brackets can now be constructed.
\newpage
A quantity $R(q,\Pi)$ is first class \cite{bi:HRT} if
\be
\vb+{+ R,\ph_{\io} \vb+}+ \we 0,\;
\io=0,1,
\label{eq:fc}
\ee
otherwise it is second class.   The second class constraints encapture 
the way in which there are more variables describing the sytem than 
is necessary,   they give rise to the Dirac matrix $C_{\io \ka}$ ,
c.f. \cite{bi:HRT} page 10,  defined by
\be
C_{\io \ka}\equiv \vb+{+ \ph_{\io},\ph_{\ka}\vb+}+.
\ee
In the present case it is given by
\ber
C_{\io \ka}&=&\Pi^{\si}\times
\left(\begin{array}{c}
         0 ~~~  -1            \\
         1 ~~~   0
\end{array}   \right)
=-i\si^{2}\Pi^{\si},\nonumber\\
C^{-1}_{\io \ka}&=&-\fr{1}{\Pi^{\si}}\times
\left(\begin{array}{c}
         0 ~~~  -1            \\
         1 ~~~   0
\end{array}   \right)
=+i\fr{\si^{2}}{\Pi^{\si}},
\ear
where $C^{-1}_{\io \ka}$ is the inverse of $C_{\io \ka}$ 
and $\si^{2}$ is a Pauli matrix, 
Bjorken and Drell (1965) \cite{bi:BD} page 378,
\ber
\si^2=
\left(\begin{array}{c}
    0~~~~-i  \\
    +i~~~~0 
\end{array}   \right).
\ear
The Dirac bracket is defined by
c.f. \cite{bi:HRT} page 11
\be
\vb+{+ A,B \vb+}+*
\equiv\vb+{+ A,B \vb+}+
-\vb+{+ A,\ph_{\io} \vb+}+
C^{-1}_{\io \ka}\vb+{+ \ph_{\ka},B \vb+}+.
\label{eq:db}
\ee
In the present case this gives the Dirac bracket
\ber
\verb+{+ A,B \verb+}+*
=\verb+{+ A,B \verb+}+
&+&\fr{1}{\Pi^{\si}}\fr{\de B}{\de \th}\left(
\fr{\de A}{\de s}-\th\fr{\de A}{\de \si}
+\Pi^{\si}\fr{\de A}{\de \Pi^{\th}}\right)\nonumber\\
&-&\fr{1}{\Pi^{\si}}\fr{\de A}{\de \th}\left(
\fr{\de B}{\de s}-\th\fr{\de B}{\de \si}
+\Pi^{\si}\fr{\de B}{\de \Pi^{\th}}\right).
\label{eq:pdb}
\ear
Consistency is checked by noting
\ber
H_{\la}&=&H_{o}-\vb+{+ H_{o},\ph^{\io} \vb+}+
                                                C^{-1}_{\io \ka},\nonumber\\
\la_{\ka}&=&-\vb+{+ H_{o},\ph^{\io} \vb+}+
                                                C^{-1}_{\io \ka}.
\label{eq:consischeck}
\ear
from which $H_{\la}$ given by 
\ref{eq:conham} can be recovered with the correct $\lambda$ 's.
\newpage
\section{Quantization.}
\label{quantization}
To quantize a classical dynamical system the Dirac bracket is replaced 
by the commutator
\be
\vb+{+ A,B \vb+}+*
\rightarrow
\fr{1}{i\hb}
\left[ \hat{A}\hat{B}-\hat{B}\hat{A} \right],
\label{eq:diracquant}
\ee
where $\hb$ is Planck's constant divided by $2\pi$ 
the hat on $\hat{A}$ and $\hat{B}$ signifies that 
the variables are now operators,
from now on the hat is dropped as the ellipis is immediate.
There are various correspondence criteria 
which one would hope to investigate,
for example:   as $\hb\rightarrow 0 $  there should be 
{\it firstly} the same time evolution,
{\it secondly} the same stress,
{\it thirdly}  the first law should be recovered.
Another correspondence criteria can be called the 
{\it particle number criteria}:  
the particle number n (see equation \ref{eq:pnh})
should bear a relation to the quantum 
particle number constructed from creation and destruction operators.   
An intermediate aim,  between formal quantization achieved by replacing 
field and momenta Dirac brackets with commutators,   and establishing contact 
with applications is to produce a quantum perfect fluid.   This could be 
obtained from brackets involving the numbered field,  the angular momentum 
and so on,  or from brackets involving a mixture of these and geometric 
objects.   However no progress has been made so far in finding a quantum 
perfect fluid,  so that attention is restricted to the
implications of replacing 
brackets consisting solely of individual components of fields and momenta with
commutators.   Effecting the replacement of the 15 Dirac brackets between the 
fields and momenta there are 4 non-vanishing commutators
\ber
&[&\Pi^{\si} \si - \si \Pi^\si]= -i\hb,\;\;
[\Pi^{\th} \th - \th \Pi^{\th}] = 0,\;\;
[\Pi^{s} s - s \Pi{^s}= -i\hb,\nonumber\\
&[&\si \th - \th \si]= -i\hb\fr{\th}{\Pi^{\si}},\;\;
[\th s - s \th]=-i\hb\fr{1}{\pi^{\si}},\;\;
[\si s - s \si] = 0.
\label{eq:comrel}
\ear
The last two commutators have the operator $\Pi^\si$ in the denominator.
This might not be well-defined.   To avoid $\Pi^\si$ in the denominator 
we multiply by the operator $\Pi^\si$,  using the first commutation of
\ref{eq:comrel} it turns out that multiplying on the left or multiplying 
on the right are equivalent so that
\ber
\left[ \si \th 
- \th \si \right] \Pi^{\si}
=\Pi^{\si}\left[ \si \th - \th \si \right]
=-i\hb\th,\nonumber\\
\left[ \th s - s \th \right] \Pi^{\si}
=\Pi^{\si}\left[ \th s - s \th\right]
=-i\hb.
\label{eq:comrel2}
\ear
These results are in accord with the equations deduced if the Dirac brackets
$\vb+{+q^i,q^{j}\Pi^{k}\vb+}+* $  are replaced by commutators.   
Left and right multiplication by  $\Pi^{\si} $ are also equivalent 
if anti-commutation rather than commutation is considered. 

The quantum Hamiltonian is
\ber
H_{q}=
&+&l_{1}\Pi^{\si}\dot{\si}+l_{2}\dot{\si}\Pi^{\si}
-l_{3}\Pi^{\si}\th\dot{s}-l_{4}\Pi^{\si}\dot{s}\th    \nonumber\\
&-&l_{5}\Pi^{\si}\dot{s}-l_{6}\th\dot{s}\Pi^{\si}
-l_{7}\dot{s}\Pi^{\si}\th-l_{8}\dot{s}\th\Pi^{\th}-p
\label{eq:qh}
\ear
where the l's are constant and obey
$l_{1}+l_{2}=1$ and $l_{3}+l_{4}+l_{5}+l_{6}+l_{7}+l_{8}=1$.
using the commutation relations \ref{eq:comrel2}
the quantum Hamiltonian \ref{eq:qh} becomes
\be
H_{q}=\Pi^{\si}(\dot{\si}-\th \dot{s})-i\hb\Th l-p,
\label{eq:qh2}
\ee
where $l=l_{2}+l_{4}+l_{7}+l_{8}$ is called the ordering constant:
it is of undefined size but is can be taken to be of order unity.   
Because the Dirac bracket of {\it p} with anything vanishes the commutators 
with {\it p} also vanish and {\it p} can be taken to be p{\it 1},
where {\it 1} is the identity element.
To investigate the algebraic implications of 
\ref{eq:comrel2} and \ref{eq:qh2}
label the 6 operators  by v's,
\be
v_{1}=\si,\;
v_{2}= s,\;
v_{3}=\th,\;
v_{4}=\Pi^{\si},\;
v_{5}=\Pi^{s},
v_{6}=\Pi^{\th}.
\label{eq:la}
\ee
$v_{6}$ commutes with everything and can be disregarded.   Of the remaining 
commutators only 4 are non-zero.   In units $\hb=1$,
\ref{eq:comrel2}  and \ref{eq:la}
give the algebra
\ber
v_{4}(v_{3}v_{2}-v_{2}v_{3})&=&-i,\nonumber\\
v_{4}(v_{1}v_{3}-v_{3}v_{1})&=&-iv_{3},\nonumber\\
v_{4}v_{1}-v_{1}v_{4}&=&-i,\nonumber\\
v_{5}v_{2}-v_{2}v_{5}&=&-i.
\ear
This algebra does not seem to be realizable in terms of matrices and
differential operators,  the closest algebras are found in \cite{bi:OK}.    
If a commutator is constructed with a time derivative of the field or 
momenta,  the same algebra results but multiplied by a term in the expansion.
Similarly if m time derivatives occur,  the algebra is multiplied by the 
expansion to the power of m.
\section{Acknowledgements}
I would like to thank Prof.T.W.B.Kibble for discussion of some of the 
points that occur.   This work was supported in part by the South
African Foundation for Research and Development (FDA).

\end{document}